# Dislocations and cracks in generalized continua

Markus Lazar

## Synonyms

Generalized Continua, Higher Order Continua, Gradient Elasticity, Nonlocality, Defects, Dislocations, Cracks, Non-singular Dislocation Continuum Theory, Dislocation Core, Regularization, Eigenstrain.

## Definitions

Dislocations play a key role in the understanding of many phenomena in solid state physics, materials science, crystallography and engineering. Dislocations are line defects producing distortions and self-stresses in an otherwise perfect crystal lattice. In particular, dislocations are the primary carrier of crystal plasticity and in dislocation based fracture mechanics.

Using classical continuum theories, the fields produced by defects (dislocations, disclinations and cracks) possess singularities since classical continuum theories are not valid near the defect (Kröner 1958; de Wit 1973). Singularities and infinities are an important problem in classical continuum theories. In order to obtain singularity-free fields, a proper regularization method must be used.

On the other hand, generalized continuum theories (e.g. micropolar elasticity, couple stress elasticity, micromorphic elasticity, nonlocal elasticity, gradient elasticity) are generalizations of classical elasticity theory and contain internal length scales (Mindlin 1964; Eringen 1999, 2002). Which of these theories is a proper theory delivering singularity-free fields of defects? Micropolar elasticity and couple stress elasticity give additional (non-classical) singularities in addition to the clas-

Markus Lazar
Department of Physics, Darmstadt University of Technology, Hochschulstr. 6, D-64289 Darmstadt, Germany, e-mail: `lazar@fkp.tu-darmstadt.de`

sical ones (Eringen 1999). Nonlocal elasticity regularizes only the stress fields and the strain and displacement fields remain singular (Eringen 2002). Strain gradient elasticity is able to deliver singularity-free fields of the displacement, elastic distortion and stress of defects (Gutkin and Aifantis 1996, 1997, 1999; Lazar and Maugin 2005, 2006; Lazar et al. 2005; Lazar 2017).

## Background

Strain gradient elasticity is a generalization of linear elasticity which includes elastic strain gradient terms to account microstructural effects and characteristic internal lengths (Mindlin 1964). Moreover, strain gradient elasticity is a proper theory in order to model defects without (unphysical) singularities (Gutkin and Aifantis 1999; Lazar and Maugin 2006). In fact, strain gradient elasticity gives a regularization based on higher order partial differential equations with non-singular Green functions, it gives a dislocation core spreading, and the size and width of a dislocation (Lazar 2014). Strain gradient elasticity is able to describe size effects since it is a theory containing characteristic length scales. Weak nonlocality is relevant due to discreteness of crystals. Gradient elasticity contains hyperstresses with information of microstructure at small scales and the relation to atomistic structure of higher order interatomic interactions. In this sense, gradient elasticity theory represents the bridge between elasticity theory and atomistic theories. It is noticed that the non-singular analytical solutions of defects have an advantage over numerical solutions (like FEM simulations and molecular statics), especially in new areas of research where benchmark solutions do not exist. In particular, strain gradient elasticity theory delivers a singularity-free and parameter-free continuum theory (Po et al. 2017).

This article deals with strain gradient elasticity of defects. The presented version of strain gradient elasticity gives a non-singular dislocation continuum theory.

## Strain gradient elasticity

In Mindlin's strain gradient elasticity of form II (Mindlin 1964; Mindlin and Eshel 1968), the strain energy density reads for centrosymmetric materials

$$W = \frac{1}{2} C_{ijkl} e_{ij} e_{kl} + \frac{1}{2} D_{ijmkln} \partial_m e_{ij} \partial_n e_{kl} \,, \tag{1}$$

where $e_{ij} = 1/2(\beta_{ij} + \beta_{ji})$ is the elastic strain tensor, $\beta_{ij}$ is the elastic distortion tensor and $\partial_m = \partial/\partial x_m$ denotes the partial derivative. The tensors $C_{ijkl}$ and $D_{ijmkln}$ are constitutive tensors of strain gradient elasticity. Using the constitutive assumption (Lazar and Kirchner 2007; Lazar 2016)

$$D_{ijmkln} = \ell^2 \delta_{mn} C_{ijkl} \,, \tag{2}$$

where $\delta_{mn}$ is the Kronecker delta tensor and $\ell$ is a (positive) characteristic internal length of strain gradient elasticity, the strain energy density (1) reduces to (Lazar and Maugin 2005; Lazar 2013, 2014; Polizzotto 2017)

$$W = \frac{1}{2} C_{ijkl} e_{ij} e_{kl} + \frac{1}{2} \ell^2 C_{ijkl} \partial_m e_{ij} \partial_m e_{kl} \,. \tag{3}$$

For isotropic materials, the tensor of elastic moduli $C_{ijkl}$ reads

$$C_{ijkl} = \lambda \, \delta_{ij} \delta_{kl} + \mu \left( \delta_{ik} \delta_{jl} + \delta_{il} \delta_{jk} \right) \,, \tag{4}$$

where $\mu$ and $\lambda$ are the Lamé moduli. Thus, for isotropic materials strain gradient elasticity of form (3) contains 3 material parameters, namely the 2 Lamé moduli ($\mu$, $\lambda$) and 1 characteristic internal length ($\ell$).

The Cauchy stress tensor is given by

$$\sigma_{ij} = \frac{\partial W}{\partial e_{ij}} = C_{ijkl} e_{kl} \,, \tag{5}$$

and the double stress tensor is given by

$$\tau_{ijk} = \frac{\partial W}{\partial (\partial_k e_{ij})} = \ell^2 C_{ijmn} \partial_k e_{mn} = \ell^2 \partial_k \sigma_{ij} \,. \tag{6}$$

The equilibrium condition (for vanishing body forces) reads in terms of the Cauchy stress tensor and double stress tensor

$$\partial_j (\sigma_{ij} - \partial_k \tau_{ijk}) = 0 \,. \tag{7}$$

Using Eq. (6), Eq. (7) reads

$$\partial_j (1 - \ell^2 \Delta) \sigma_{ij} = 0 \,, \tag{8}$$

where $\Delta$ is the Laplace operator.

In presence of defects such as dislocations, the total distortion tensor $\beta_{ij}^{\mathrm{T}}$ can be decomposed into an elastic distortion tensor $\beta_{ij}$ and a plastic distortion tensor $\beta_{ij}^{\mathrm{P}}$

$$\beta_{ij}^{\mathrm{T}} = \partial_j u_i = \beta_{ij} + \beta_{ij}^{\mathrm{P}} \,, \tag{9}$$

where $u_i$ denotes the displacement vector. Since dislocations cause self-stresses, body forces are zero. The dislocation density tensor is defined in terms of the plastic and elastic distortion tensors (Kröner 1958)

$$\alpha_{ij} = -\varepsilon_{jkl} \partial_k \beta_{il}^{\mathrm{P}} \,, \quad \text{or} \quad \alpha_{ij} = \varepsilon_{jkl} \partial_k \beta_{il} \,, \tag{10}$$

where $\varepsilon_{jkl}$ denotes the Levi-Civita tensor. The dislocation density tensor satisfies the dislocation Bianchi identity

$$\partial_j \alpha_{ij} = 0,$$ (11)

which means that a dislocation cannot end inside the medium.

The equilibrium condition (8) reads in terms of the displacement vector and the plastic distortion tensor

$$LL_{ik}u_k = C_{ijkl}\partial_j L\beta_{kl}^{\mathrm{P}},$$ (12)

where

$$L = 1 - \ell^2 \Delta$$ (13)

is the (modified) Helmholtz operator, and

$$L_{ik} = C_{ijkl}\partial_j\partial_l$$ (14)

is the Navier operator. Using two inhomogeneous Helmholtz equations (Ru and Aifantis 1993; Lazar 2014)

$$Lu_k = u_k^0,$$ (15)

$$L\beta_{kl}^{\mathrm{P}} = \beta_{kl}^{\mathrm{P},0},$$ (16)

an inhomogeneous Navier equation known from classical eigenstrain theory (Mura 1987) is obtained

$$L_{ik}u_k^0 = C_{ijkl}\partial_j\beta_{kl}^{\mathrm{P},0}.$$ (17)

Therefore, the fields $u_k^0$ and $\beta_{kl}^{0}$ may be identified with the classical displacement vector and classical plastic distortion tensor of classical incompatible elasticity theory. Using Eqs. (12) and (16), the displacement vector $u_k$ is determined by an inhomogeneous Helmholtz-Navier equation (Lazar 2013, 2014)

$$LL_{ik}u_k = C_{ijkl}\partial_j\beta_{kl}^{\mathrm{P},0},$$ (18)

where the right hand side is given by the gradient of the classical plastic distortion tensor (classical eigendistortion).

Eq. (12) can be written in terms of the elastic distortion tensor and the dislocation density tensor

$$LL_{ik}\beta_{km} = -C_{ijkl}\varepsilon_{mlr}\partial_j L\alpha_{kr}.$$ (19)

Using two inhomogeneous Helmholtz equations (Lazar 2014)

**Table 1** Calculated characteristic lengths and equilibrium lattice parameter for several fcc and bcc crystals via ab initio (Shodja et al. 2013).

| Material | Crystal | $\ell_1$ (Å) | $\ell_2$ (Å) | $\ell = \dfrac{\ell_1 + \ell_2}{2}$ (Å) | $a$ (Å) | $\ell/a$ |
|----------|---------|--------------|--------------|------------------------------------------|---------|----------|
| Ir | fcc | 2.1523 | 1.8217 | 1.9870 | 3.87 | 0.51 |
| Pt | fcc | 2.4480 | 1.6353 | 2.0416 | 3.92 | 0.52 |
| Al | fcc | 2.3415 | 1.6582 | 1.9998 | 4.05 | 0.49 |
| W  | bcc | 1.6460 | 2.2026 | 1.9243 | 3.15 | 0.61 |
| V  | bcc | 1.5519 | 2.1710 | 1.8614 | 3.02 | 0.62 |
| Mo | bcc | 1.6380 | 2.2438 | 1.9409 | 3.16 | 0.61 |

$$L\beta_{km} = \beta_{km}^0, \tag{20}$$

$$L\alpha_{kr} = \alpha_{kr}^0, \tag{21}$$

an inhomogeneous Navier equation known from classical dislocation theory (Mura 1987) is obtained

$$L_{ik}\beta_{km}^0 = -C_{ijkl}\varepsilon_{mlr}\partial_j\alpha_{kr}^0. \tag{22}$$

Therefore, the fields $\beta_{km}^0$ and $\alpha_{kr}^0$ may be identified with the classical elastic distortion tensor and classical dislocation density tensor of classical incompatible elasticity theory of dislocations. Using Eqs. (19) and (21), the elastic distortion tensor $\beta_{km}$ satisfies an inhomogeneous Helmholtz-Navier equation (Lazar 2013, 2014)

$$LL_{ik}\beta_{km} = -C_{ijkl}\varepsilon_{mlr}\partial_j\alpha_{kr}^0, \tag{23}$$

where the right hand side is given by the gradient of the classical dislocation density tensor $\alpha_{kr}^0$.

An important issue in strain gradient elasticity is the determination of the characteristic lengths in addition to the elastic constants. In Table 1, the characteristic internal lengths $\ell_1$ and $\ell_2$ of Mindlin's isotropic strain gradient elasticity theory (Mindlin 1964) are given for several fcc and bcc crystals determined by ab initio density functional theory (DFT) method (Shodja et al. 2013). From Mindlin's characteristic lengths $\ell_1$ and $\ell_2$, the characteristic length of (simplified) strain gradient elasticity can be determined (see Table 1)

$$\ell = \frac{\ell_1 + \ell_2}{2}. \tag{24}$$

The characteristic length $\ell$ is the average of the two characteristic lengths $\ell_1$ and $\ell_2$ of Mindlin's isotropic gradient elasticity theory (Shodja and Tehranchi 2010). A few cubic crystals such as tungsten (W) and aluminum (Al) are elastically isotropic or nearly isotropic materials (Dederichs and Leibfried 1969). Tungsten (W) is the best material to test the theory of isotropic strain gradient elasticity.

## Screw dislocation

### *Classical solution*

Consider a Volterra screw dislocation at position $(x, y) = (0, 0)$ whose Burgers vector $b_z$ and dislocation line coincide with the direction of the $z$-axis of a Cartesian coordinate system. In the framework of classical incompatible elasticity, the discontinuous displacement field with the branch cut at $x = 0$ and for $y < 0$ reads (Leibfried and Dietze 1949)

$$u_z^0 = -\frac{b_z}{2\pi} \arctan \frac{x}{y}. \tag{25}$$

The total distortion, consisting of the incompatible elastic distortion and the incompatible plastic distortion, is the gradient of the displacement (25) and has two non-vanishing components

$$\beta_{zx}^{\mathrm{T},0} = \partial_x u_z^0 = -\frac{b_z}{2\pi} \left( \frac{y}{r^2} + 2\pi\, \delta(x) H(-y) \right), \tag{26}$$

$$\beta_{zy}^{\mathrm{T},0} = \partial_y u_z^0 = \frac{b_z}{2\pi} \frac{x}{r^2}, \tag{27}$$

where $r = \sqrt{x^2 + y^2}$. Here $\delta(x)$ denotes the one-dimensional Dirac delta function and $H(y)$ is the Heaviside step function defined by

$$H(y) = \begin{cases} 0, & y < 0, \\ 1, & y > 0. \end{cases} \tag{28}$$

The elastic distortion and stress fields of the Volterra screw dislocation read (de Wit 1973)

$$\beta_{zx}^0 = \frac{\sigma_{zx}^0}{\mu} = -\frac{b_z}{2\pi} \frac{y}{r^2}, \tag{29}$$

$$\beta_{zy}^0 = \frac{\sigma_{zy}^0}{\mu} = \frac{b_z}{2\pi} \frac{x}{r^2}. \tag{30}$$

These fields are singular at the dislocation line $(x = 0, y = 0)$.

Because the last term in Eq. (26) is discontinuous and singular at the branch cut, the plastic distortion reads

$$\beta_{zx}^{\mathrm{P},0} = -b_z\, \delta(x) H(-y). \tag{31}$$

The plastic distortion gives rise to the dislocation density of the screw dislocation

$$\alpha_{zz}^0 = \partial_y \beta_{zx}^{\mathrm{P},0} = b_z\, \delta(x)\delta(y), \tag{32}$$

in terms of Dirac delta functions.

### *Gradient solution*

Using Fourier transform (Lazar and Maugin 2006), the displacement field of a screw dislocation, being solution of Eq. (15), reads

$$u_z = \frac{b_z}{2\pi} w(x, y),$$ (33)

where

$$w = -\arctan\frac{x}{y} + \int_0^\infty \frac{s\sin(sx)}{s^2 + \frac{1}{\ell^2}} \left[ \operatorname{sgn}(y) \, e^{-|y|\sqrt{s^2 + \frac{1}{\ell^2}}} + 2H(-y) \right] ds$$ (34)

and $\operatorname{sgn}(y)$ denotes the signum function defined by

$$\operatorname{sgn}(y) = \begin{cases} -1, & y < 0, \\ 1, & y > 0. \end{cases}$$ (35)

When $y \to 0^+$, the displacement field (34) reduces to

$$w(x, 0^+) = -\frac{\pi}{2} \operatorname{sgn}(x) \left\{ 1 - e^{-|x|/\ell} \right\}.$$ (36)

The gradient term appearing in Eq. (36) leads to a smoothing of the displacement profile unlike the jump occurring in the classical solution. The smoothing depends on the length scale $\ell$.

For the elastic distortion tensor, solution of Eq. (20) gives (Lazar and Maugin 2006)

$$\beta_{zx} = -\frac{b_z}{2\pi} \frac{y}{r^2} \left\{ 1 - \frac{r}{\ell} K_1(r/\ell) \right\},$$ (37)

$$\beta_{zy} = \frac{b_z}{2\pi} \frac{x}{r^2} \left\{ 1 - \frac{r}{\ell} K_1(r/\ell) \right\}.$$ (38)

The corresponding stress components of a screw dislocation read (Lazar and Maugin 2005; Gutkin and Aifantis 1999)

$$\sigma_{zx} = -\frac{\mu b_z}{2\pi} \frac{y}{r^2} \left\{ 1 - \frac{r}{\ell} K_1(r/\ell) \right\},$$ (39)

$$\sigma_{zy} = \frac{\mu b_z}{2\pi} \frac{x}{r^2} \left\{ 1 - \frac{r}{\ell} K_1(r/\ell) \right\},$$ (40)

where $K_n$ denotes the modified Bessel function of order $n$. The appearance of the modified Bessel function $K_1$ in Eqs. (37)–(40) leads to the regularization of the classical singularity of order $\mathcal{O}(1/r)$ at the dislocation line and gives non-singular stresses and non-singular elastic distortions. The non-singular stresses are zero at the dislocation line. The stress $\sigma_{zy}$ has its extreme value $|\sigma_{zy}(x,0)| \simeq 0.399 \frac{\mu b_z}{2\pi \ell}$ at $|x| \simeq 1.114\ell$, whereas the stress $\sigma_{zx}$ has its extreme value $|\sigma_{zx}(0,y)| \simeq 0.399 \frac{\mu b_z}{2\pi \ell}$ at $|y| \simeq 1.114\ell$. For $W$ with $\ell = 0.61a$: $\sigma_{zy}$ has its extreme value $|\sigma_{zy}(x,0)| \simeq 0.654 \frac{\mu b_z}{2\pi a}$ at $|x| \simeq 0.68a$, whereas $\sigma_{zx}$ has its extreme value $|\sigma_{zx}(0,y)| \simeq 0.654 \frac{\mu b_z}{2\pi a}$ at $|y| \simeq 0.68a$.

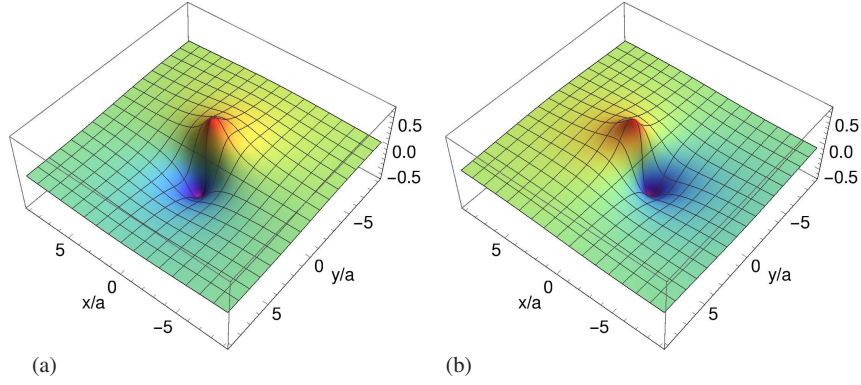

**Fig. 1** Stress fields of a screw dislocation in strain gradient elasticity for $\ell = 0.61a$ near the dislocation line: (a) $\sigma_{zx}$ and (b) $\sigma_{zy}$ are given in units of $\mu b_z/[2\pi a]$ .

For the plastic distortion tensor, solution of Eq. (16) reads

$$\beta_{zx}^{\mathrm{P}} = -\frac{b_z}{2\pi} \int_0^\infty \frac{\cos(sx)}{1+\ell^2 s^2} \left[ \mathrm{sgn}(y)\, \mathrm{e}^{-|y|\sqrt{s^2+\frac{1}{\ell^2}}} + 2H(-y) \right] \mathrm{d}s. \qquad (41)$$

When $y \to 0$, the plastic distortion (41) reduces to

$$\beta_{zx}^{\mathrm{P}}(x,0) = -\frac{b_z}{2\pi} \frac{\pi}{2} \frac{1}{\ell} \mathrm{e}^{-|x|/\ell}. \qquad (42)$$

Unlike the classical plastic distortion which is singular at $x = 0$ due to $\delta(x)$ in Eq. (31), Eq. (42) is smooth. For the dislocation density of a screw dislocation, the solution of Eq. (21) is given by (Lazar and Maugin 2006; Lazar et al. 2005)

$$\alpha_{zz} = \frac{b_z}{2\pi \ell^2} K_0(r/\ell), \qquad (43)$$

which possesses a logarithmic singularity at the dislocation line, but it is smoother than the classical dislocation density tensor (32). The dislocation density tensor defines the dislocation core region and determines the shape and size of the dislocation core. Thus $\alpha_{ij}$ has the physical meaning of a dislocation core tensor. Eq. (43) obtained in strain gradient elasticity describes a dislocation core spreading (see Fig.

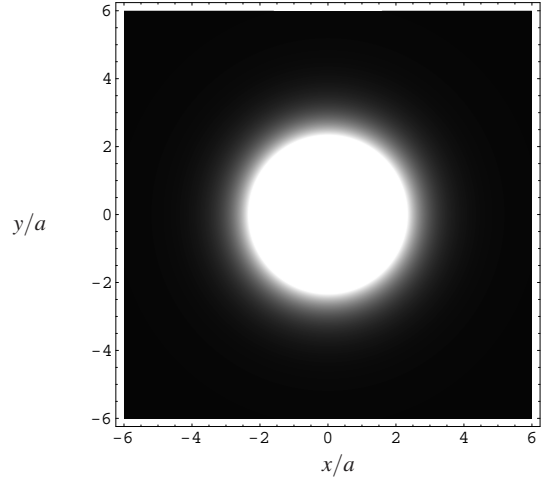

**Fig. 2** Contour of the dislocation density of a screw dislocation in strain gradient elasticity for $\ell = 0.61a$.

2) and represents the weak nonlocality present in the dislocation core region. Fig. 2 shows that the dislocation core radius is obtained as $2.5a \leq r_c \leq 3a$ for W.

## Edge dislocation

### *Classical solution*

Consider a Volterra edge dislocation at position $(x,y) = (0,0)$ whose Burgers vector $b_x$ is parallel to the $x$-axis and the dislocation line coincides with the $z$-axis of the Cartesian coordinate system. The classical discontinuous displacements with the branch cut at $x = 0$ and for $y < 0$ are (Leibfried and Lücke 1949; Seeger 1955)

$$u_x^0 = -\frac{b_x}{2\pi}\left(\arctan\frac{x}{y} - \frac{xy}{2(1-\nu)r^2}\right),\tag{44}$$

$$u_y^0 = -\frac{b_x}{4\pi(1-\nu)}\left((1-2\nu)\ln r + \frac{x^2}{r^2}\right),\tag{45}$$

where $\nu$ is the Poisson ratio.

The non-vanishing components of the elastic distortion read (de Wit 1973)

$$\beta_{xx}^0 = -\frac{b_x}{4\pi(1-\nu)}\frac{y}{r^2}\left\{(1-2\nu)+\frac{2x^2}{r^2}\right\}, \tag{46}$$

$$\beta_{xy}^0 = \frac{b_x}{4\pi(1-\nu)}\frac{x}{r^2}\left\{(3-2\nu)-\frac{2y^2}{r^2}\right\}, \tag{47}$$

$$\beta_{yx}^0 = -\frac{b_x}{4\pi(1-\nu)}\frac{x}{r^2}\left\{(1-2\nu)+\frac{2y^2}{r^2}\right\}, \tag{48}$$

$$\beta_{yy}^0 = -\frac{b_x}{4\pi(1-\nu)}\frac{y}{r^2}\left\{(1-2\nu)-\frac{2x^2}{r^2}\right\}, \tag{49}$$

and the stress components are

$$\sigma_{xx}^0 = -\frac{\mu b_x}{2\pi(1-\nu)}\frac{y}{r^4}\left(y^2+3x^2\right), \tag{50}$$

$$\sigma_{yy}^0 = -\frac{\mu b_x}{2\pi(1-\nu)}\frac{y}{r^4}\left(y^2-x^2\right), \tag{51}$$

$$\sigma_{xy}^0 = \frac{\mu b_x}{2\pi(1-\nu)}\frac{x}{r^4}\left(x^2-y^2\right), \tag{52}$$

$$\sigma_{zz}^0 = -\frac{\mu b_x \nu}{\pi(1-\nu)}\frac{y}{r^2}. \tag{53}$$

These fields are singular at the dislocation line $(x=0, y=0)$.

The plastic distortion reads

$$\beta_{xx}^{0,\mathrm{P}} = -b_x\,\delta(x)H(-y), \tag{54}$$

and the dislocation density of a single edge dislocation located at the position $(0,0)$ has the non-vanishing component

$$\alpha_{xz}^0 = \partial_y \beta_{xx}^{0,\mathrm{P}} = b_x\,\delta(x)\delta(y). \tag{55}$$

### Gradient solution

Using Eq. (15), the displacement fields of an edge dislocation are (Lazar and Maugin 2006)

$$u_x = \frac{b_x}{4\pi(1-\nu)}\left\{2(1-\nu)w(x,y)+\frac{xy}{r^2}-4\ell^2\frac{xy}{r^4}+\frac{2xy}{r^2}K_2(r/\ell)\right\}, \tag{56}$$

$$u_y = -\frac{b_x}{4\pi(1-\nu)}\left\{(1-2\nu)\big(\ln r + K_0(r/\ell)\big)+\frac{x^2}{r^2}-2\ell^2\frac{x^2-y^2}{r^4}+\frac{x^2-y^2}{r^2}K_2(r/\ell)\right\}, \tag{57}$$

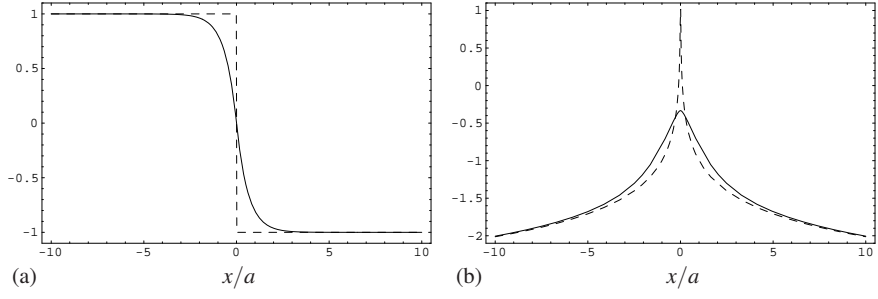

**Fig. 3** Displacement fields of an edge dislocation in strain gradient elasticity for $\ell = 0.61a$ (solid line) and Volterra model (dashed line): a) $u_x(x, 0^+)$ is given in units of $b_x/4$, and b) $u_y(x, 0)$ is given in units of $b_x/[4\pi(1-\nu)]$ for $\nu = 0.28$.

where the displacement profile function $w(x, y)$ is given by Eq. (34). The displacement fields (56) and (57) are plotted in Fig. 3. Both displacement fields are non-singular and the field (56) possesses a discontinuity due to the dislocation profile function $w(x, y)$.

Using Eq. (20), the elastic distortions of an edge dislocation read (Lazar and Maugin 2006; Lazar 2013)

$$\beta_{xx} = -\frac{b_x}{4\pi(1-\nu)} \frac{y}{r^2} \left\{ (1-2\nu) + \frac{2x^2}{r^2} + \frac{4\ell^2}{r^4}(y^2 - 3x^2) - \frac{2(y^2 - 3x^2)}{r^2} K_2(r/\ell) \right. $$
$$\left. - \frac{2(y^2 - \nu r^2)}{\ell r} K_1(r/\ell) \right\}, \quad (58)$$

$$\beta_{xy} = \frac{b_x}{4\pi(1-\nu)} \frac{x}{r^2} \left\{ (3-2\nu) - \frac{2y^2}{r^2} - \frac{4\ell^2}{r^4}(x^2 - 3y^2) + \frac{2(x^2 - 3y^2)}{r^2} K_2(r/\ell) \right. $$
$$\left. - \frac{2(y^2 + (1-\nu)r^2)}{\ell r} K_1(r/\ell) \right\}, \quad (59)$$

$$\beta_{yx} = -\frac{b_x}{4\pi(1-\nu)} \frac{x}{r^2} \left\{ (1-2\nu) + \frac{2y^2}{r^2} + \frac{4\ell^2}{r^4}(x^2 - 3y^2) - \frac{2(x^2 - 3y^2)}{r^2} K_2(r/\ell) \right. $$
$$\left. + \frac{2(y^2 - (1-\nu)r^2)}{\ell r} K_1(r/\ell) \right\}, \quad (60)$$

$$\beta_{yy} = -\frac{b_x}{4\pi(1-\nu)} \frac{y}{r^2} \left\{ (1-2\nu) - \frac{2x^2}{r^2} - \frac{4\ell^2}{r^4}(y^2 - 3x^2) + \frac{2(y^2 - 3x^2)}{r^2} K_2(r/\ell) \right. $$
$$\left. - \frac{2(x^2 - \nu r^2)}{\ell r} K_1(r/\ell) \right\}. \quad (61)$$

The elastic distortions (58)–(61) are non-singular.

The stresses read (Lazar and Maugin 2005; Gutkin and Aifantis 1999)

$$\sigma_{xx} = -\frac{\mu b_x}{2\pi(1-\nu)}\frac{y}{r^4}\Big\{\big(y^2+3x^2\big)+\frac{4\ell^2}{r^2}\big(y^2-3x^2\big)-2y^2\frac{r}{\ell}K_1(r/\ell)$$
$$-2\big(y^2-3x^2\big)K_2(r/\ell)\Big\}, \quad (62)$$

$$\sigma_{yy} = -\frac{\mu b_x}{2\pi(1-\nu)}\frac{y}{r^4}\Big\{\big(y^2-x^2\big)-\frac{4\ell^2}{r^2}\big(y^2-3x^2\big)-2x^2\frac{r}{\ell}K_1(r/\ell)$$
$$+2\big(y^2-3x^2\big)K_2(r/\ell)\Big\}, \quad (63)$$

$$\sigma_{xy} = \frac{\mu b_x}{2\pi(1-\nu)}\frac{x}{r^4}\Big\{\big(x^2-y^2\big)-\frac{4\ell^2}{r^2}\big(x^2-3y^2\big)-2y^2\frac{r}{\ell}K_1(r/\ell)$$
$$+2\big(x^2-3y^2\big)K_2(r/\ell)\Big\}, \quad (64)$$

$$\sigma_{zz} = -\frac{\mu b_x\nu}{\pi(1-\nu)}\frac{y}{r^2}\Big\{1-\frac{r}{\ell}K_1(r/\ell)\Big\}. \quad (65)$$

The stresses (62)–(65) are non-singular and continuous. They are zero at the dislocation line. In fact, the "classical" singularities are eliminated. The stresses (62)–(65) have the following extreme values: $|\sigma_{xx}(0,y)| \simeq 0.546\frac{\mu b_x}{2\pi(1-\nu)\ell}$ at $|y| \simeq 0.996\ell$, $|\sigma_{yy}(0,y)| \simeq 0.260\frac{\mu b_x}{2\pi(1-\nu)\ell}$ at $|y| \simeq 1.494\ell$, $|\sigma_{xy}(x,0)| \simeq 0.260\frac{\mu b_x}{2\pi(1-\nu)\ell}$ at $|x| \simeq 1.494\ell$, and $|\sigma_{zz}(0,y)| \simeq 0.399\frac{\mu b_x\nu}{\pi(1-\nu)\ell}$ at $|y| \simeq 1.114\ell$. For W with $\ell = 0.61a$: $|\sigma_{xx}(0,y)| \simeq 0.895\frac{\mu b_x}{2\pi(1-\nu)a}$ at $|y| \simeq 0.61a$, $|\sigma_{yy}(0,y)| \simeq 0.426\frac{\mu b_x}{2\pi(1-\nu)a}$ at $|y| \simeq 0.91a$, $|\sigma_{xy}(x,0)| \simeq 0.426\frac{\mu b_x}{2\pi(1-\nu)a}$ at $|x| \simeq 0.91a$, and $|\sigma_{zz}(0,y)| \simeq 0.654\frac{\mu b_x\nu}{\pi(1-\nu)a}$ at $|y| \simeq 0.68a$. The non-singular stresses are plotted in Fig. 4.

The plastic distortion of an edge dislocation in strain gradient elasticity reads

$$\beta^{\mathrm{P}}_{xx} = -\frac{b_x}{2\pi}\int_0^\infty \frac{\cos(sx)}{1+\ell^2 s^2}\Big[\mathrm{sgn}(y)\,\mathrm{e}^{-|y|\sqrt{s^2+\frac{1}{\ell^2}}}+2H(-y)\Big]\mathrm{d}s, \quad (66)$$

which is the solution of Eq. (16). The dislocation density of an edge dislocation is given by

$$\alpha_{xz} = \frac{b_x}{2\pi\ell^2}K_0(r/\ell), \quad (67)$$

which is the solution of Eq. (21). Such two-dimensional dislocation core shape is in agreement with the measurement of the dislocation core distribution by high resolution transmission electron microscopy and image processing.

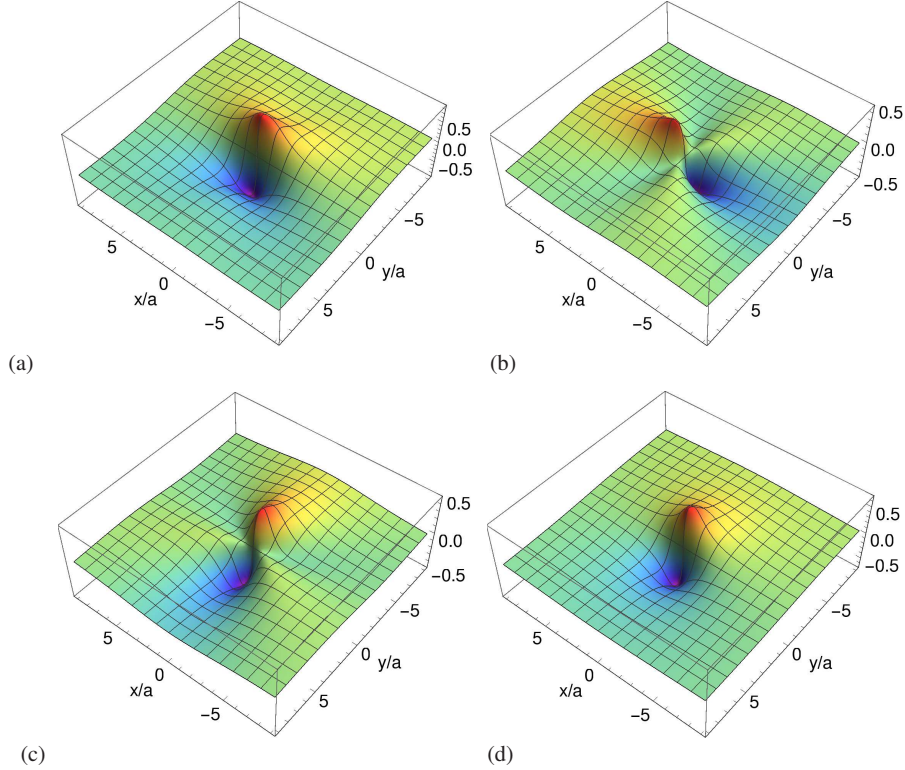

**Fig. 4** Stress fields of an edge dislocation in strain gradient elasticity for $\ell = 0.61a$: (a) $\sigma_{xx}$, (b) $\sigma_{xy}$, (c) $\sigma_{yy}$ are given in units of $\mu b_x/[2\pi(1-\nu)a]$ and (d) $\sigma_{zz}$ is given in units of $\mu b_x \nu/[\pi(1-\nu)a]$.

## Towards singularity-free cracks

Since a dislocation is the building block of a crack, the crack fields can be calculated based on the dislocation solutions in the framework of dislocation based fracture mechanics (Weertman 1996) and distributed dislocation technique (Hills et al. 1996). The non-singular solutions of screw and edge dislocations given in subsections (screw dislocation) and (edge dislocation) can be used in order to construct non-singular solutions of crack problems. The cracks can be modeled by a continuous distribution of straight dislocations. The corresponding dislocation density is determined by using boundary conditions resulting from a variational formulation. Using the non-singular stress fields of a screw dislocation and (glide and climb) edge dislocations, non-singular stress fields of mode III, mode II and mode I were given by Mousavi and Lazar (2015) in the framework of nonlocal elasticity. In the framework of (simplified) strain gradient elasticity, the non-singular solution for a crack of a mode III were given by Mousavi and Aifantis (2015), and the non-singular solutions for a cracks of a mode I and mode II were given by Mousavi and Aifantis

(2016) using the corresponding non-singular solutions of screw and edge disloca-
tions. In strain gradient elasticity, the stress, elastic strain and plastic distortion fields
of cracks are non-singular and finite. The crack opening displacement is smoother
in gradient elasticity than in classical approach. A non-singular dislocation based
fracture theory represents the unification of non-singular dislocation based plastic-
ity theory and fracture mechanics.

## Singularity-free dislocation dynamics

Since dislocations produce plasticity in crystals, the non-singular dislocation solu-
tions obtained in strain gradient elasticity are the key issue for a non-singular dis-
crete dislocation dynamics. The formulation of non-singular (discrete) dislocation
dynamics based on non-singular dislocations in strain gradient elasticity is given by
Po et al. (2014, 2017).

## Conclusions

The strain gradient elasticity described by (3) delivers a robust, singularity-free and
parameter-free continuum theory of defects, and it has been employed in the use
of nanoscale short-range elastic fields of dislocations. The analytical solutions for
the displacements, elastic distortion and stress fields demonstrate the elimination
and regularization of any singularity at the dislocation line and in the dislocation
core. These non-singular dislocation solutions can be used as benchmark solutions
for FEM simulations and atomistic simulations of dislocations, and also for discrete
dislocation dynamics, dislocation based plasticity and dislocation based fracture me-
chanics.

## Cross-References

Computational Mechanics of Generalized Continua, Micromorphic Approach to
Materials with Internal Length, Non Local Theories, Strain Gradient Theories,
Strain Gradient Plasticity, Phase Field Models and Mechanics.

## Acknowledgements

The author acknowledges a grant obtained from the Deutsche Forschungsgemein-
schaft (grant number La1974/4-1).